\documentclass[aps,pre,reprint,groupedaddress,showpacs,showkeys]{revtex4-1}

\usepackage{graphicx}
\usepackage{dcolumn}
\usepackage{bm}
\usepackage{caption}
\usepackage{subcaption}
\usepackage{amsmath}
\usepackage{amssymb}
\usepackage{color}
 
\begin{document}
\title{Percolation on  Networks with Antagonistic and Dependent Interactions. }
\author{Bhushan Kotnis}
\email{bkotnis@dese.iisc.ernet.in}
\author{Joy Kuri}
\email{kuri@dese.iisc.ernet.in}
\affiliation{Indian Institute of Science, Department of Electronic Systems Engineering, Bangalore 560012, India.}
\date{\today}

\begin{abstract}
Drawing inspiration from real world interacting systems, we study a system consisting of two networks that exhibit antagonistic and dependent interactions. By antagonistic and dependent interactions, we mean, that a proportion of functional nodes in a network cause failure of nodes in the other, while failure of nodes in the other results in failure of links in the first. As opposed to interdependent networks, which can exhibit first order phase transitions, we find that the phase transitions in such networks are continuous. Our analysis shows that, compared to an isolated network, the system is more robust against random attacks. Surprisingly, we observe a region in the parameter space where the giant connected components of both networks start oscillating. Furthermore, we find that for Erdos-Renyi and scale free networks the system oscillates only when the dependency and antagonism between the two networks is very high. We believe that this study can further our understanding of real world interacting systems.
\end{abstract}



\maketitle

\section{Introduction \label{sec:Introduction}}
In recent years, research on the structure of complex interacting systems and the dynamical processes occurring on these systems have attracted a lot of attention. A majority of studies, focused on investigating the structure and properties of complex systems, model them as a single connected network, where the link between two nodes represents an interaction between two entities \cite{Cohen2000,Newman2002,Newman2003,Kotnis2013}. However, most real world systems are composed of networks which interact with one other, such as the power distribution network and the communication network. The nodes in the power distribution network depend on communication nodes (routers) for exchange of control messages, while communication nodes depend on power stations for electricity \cite{Rinaldi2001,Panzieri2008,Rosato2008}. Due to this interdependence, failure of a proportion of nodes in either network may result in complete collapse of both networks. This is confirmed by recent studies on the percolation behavior \cite{Parshani2010,Gao2011,Gao2012} of interdependent networks; percolation analysis revealed the presence of first order phase transitions, i.e., collapse of a portion of nodes in one network may lead to catastrophic collapse of the entire system. Thus, the study of such interacting networks is key in furthering our understanding of real world systems. 
\par
 Not all real world systems exhibit interdependent behavior. In this article we study one such system. Consider a botnet (computers infected with malware) launching denial of service (DoS) attacks against SCADA  (supervisory control and data acquisition) systems which control power stations. A typical denial of service attack (DoS) happens when the target is overwhelmed with service and resource requests. This prevents legitimate users from accessing the service and may even cause the target server to shutdown. A botnet is a network of computers, infected with malware, which launch DoS attacks, due to the malware, \cite{Mirkovic2004} against Internet servers, routers, or any other critical infrastructure such as SCADA  systems. Simulation studies have shown that such a distributed denial of service attack (DDoS) on SCADA  systems can result in failure of power stations \cite{Ciancamerla2013,Markovic2013}. Also, the communication links providing connectivity to the bot net may depend on the electricity supplied by the power distribution network. Thus, a DDoS attack by a botnet may lead to the failure of power stations, which in turn may cause fragmentation of the botnet, resulting in the reduction of DDoS  attacks. The decrease in the number of DDoS  attacks may allow the failed power stations to resume operation which may also cause the failed links to become operational and the cycle may continue. Clearly, such attacks have the potential to damage the entire power distribution network. Understanding the dynamics of the system under DDoS  attacks is crucial for formulating strategies to counter such attacks.  Motivated by this real world problem, we formulate a mathematical model of two interacting networks and study their robustness against such attacks. We do this by using tools from statistical mechanics, namely, percolation theory. 
 \par
 Since the nodes in the botnet launch an attack on the nodes in the power distribution network, we call this interaction `antagonistic', a term also used in \cite{Zhao2013}.  The interaction between the power distribution network and the botnet is termed as dependent, since the links of the botnet are dependent on the nodes of the power distribution network. Since the botnet antagonizes the power distribution network, we refer to the botnet as the antagonistic network and the power distribution network as the victim network.  To reiterate, nodes infected with malware in the antagonistic network cause failure of a node in the victim network, while failure of a node in the victim network may result in failure of a link in the antagonistic network. This results in negative feedback, leading to the self regulation of attacks launched by the antagonistic network. Such self regulating mechanisms abound in biological systems such as: inter-cellular and intra-cellular machinery \cite{Martinez2003}, the mammalian immune system \cite{Kim2009}, and ecological systems \cite{Brauer2011}.   
 \par
Recently, in the research community, the study of interdependent networks is slowly  gaining traction. The robustness and phase transition properties of networks consisting of two or more interdependent networks was studied in  \cite{Parshani2010,Hu2011,Jia2011,Baxter2012,Li2012,Son2012,Davide2013,Dong2013,Yanqing2013,Bashan2013}. These studies discovered that networks of  interdependent networks can exhibit first order discontinuous phase transition making them susceptible to catastrophic failure under attacks. First order phase transitions were also observed in single networks consisting of connectivity and dependency links \cite{Bashan2011,Parshani2011}.  Furthermore, unlike isolated scale free networks, it was found that it is difficult to protect interdependent networks against an attack  \cite{Huang2011,Dong2012,Byungjoon2014} by protecting high degree nodes. However, these results may not be applicable for networks exhibiting antagonistic and dependent interactions; in fact our investigations show that these results are not observed in such systems.   
\par
A system of networks with mutual antagonistic interactions, and interdependent networks with a fraction of antagonistic nodes was recently studied in \cite{Zhao2013} and \cite{Zhao2014}, respectively. In the purely antagonistic case, nodes functioning in one network cause failure of nodes in the other and vice versa. In the mixed case \cite{Zhao2014}, the two networks are interdependent with antagonistic nodes in both the networks. This is different from the system considered here, because in our case, antagonistic nodes are present in only one network. The difference may seem minor, but it has a significant effect on the phase transitions of the system.  We show that the phase transitions observed in pure and mixed antagonistic interacting networks  are very different from those observed in interacting antagonistic dependent networks.  A very thorough but accessible introduction to percolation on multilayer networks covering interdependent as well as antagonistic interactions can be found in \cite{Boccaletti2014}.
\par
In a very recent article \cite{Majdandzic2014}, researchers investigated networks which recover spontaneously after an attack. They observe a phenomenon where the mean number of active nodes undergo a phase transition. In our model, where both the networks recover due to the negative feedback, we  observe that the size of the giant connected component (GCC)  fluctuates only in a certain region of the parameter space. 
\par
Our contributions are summarized as follows. We study analytically and numerically the percolation behavior of the system. We analytically show that  unlike the first order phase transition observed in interdependent networks, the phase transition is continuous. Although the antagonistic network depends on the victim network, our results suggests that, for Erdos-Renyi  and scale free networks,  the antagonistic network always percolates, while the attacked network may fail completely.  Additionally, we show that, in comparison with an isolated network, networks with dependent and antagonistic interactions are more robust against random attacks. More importantly, we find that such a system exhibits a threshold behavior. If the dependency and antagonism are high enough, the giant connected components in the two networks oscillate, while for low dependency and antagonism the giant connected components remain stable.  Such phenomenon is neither observed in interdependent networks \cite{Gao2012}, nor in networks exhibiting antagonistic interactions \cite{Zhao2013, Zhao2014}.  
\par
The article is organized as follows. The system model is detailed in Sec. \ref{sec:Model}, analytical results are discussed in Sec. \ref{sec:Analysis} while numerical results are outlined in Sec. \ref{sec:Results} and the implications of the results are presented in Sec. \ref{sec:Discussion}.

 \section{Model \label{sec:Model}}
 The system consists of two interacting networks. Let  $A$ be the antagonistic computer network, and $B$ be the non-antagonistic power distribution network.  Let $N_a$, $N_b$ be the number of nodes in $A$ and $B$ respectively. Let $q_a$, $q_a > 0$, be the fraction of the communication links in network $A$ that depend on the power stations (nodes) in network $B$. Let $r$, $r>0$, be the fraction of nodes (computers) in network $A$ infected by malware. We assume that the compromised computers in network $A$ launch a coordinated attack on network $B$, i.e., a node in network $B$ is attacked by at most one compromised node from $A$.  More specifically, we assume that the compromised $rN_a$ number of nodes in $A$ randomly attack $rN_a$ number of nodes from network $B$ ($rN_a  \leq N_b$).  Thus, the probability that a randomly chosen node in $B$ suffers an attack is $\frac{rN_a}{N_b}$. Assuming $N_a = N_b$, this expression reduces to $r$.
 \par
 Thus, a node in network $B$ fails when it is attacked by a compromised node from network $A$, or if it is disconnected from the giant connected component. We assume that a communication link in network $A$, depends on a randomly chosen node in $B$. Thus, a link in $A$ may fail when a node, on which it depends, fails. Without an Internet connection, it is not possible to launch a DDoS attack; hence the failure of communication links may isolate the compromised nodes, resulting in cessation of attacks. Thus a compromised node is unable to launch a DDoS attack if it is not connected to the giant connected component in $A$.  The fraction of nodes in network $A$, infected with malware, ($r$), is assumed to be a constant. However, the fraction of nodes that launch an attack depends on the GCC of the network, which may change with time.
 \par
 Let $P_{a}(k)$ and $P_{b}(k)$ be the degree distribution of networks $A$ and $B$ respectively. Let $G^a(f^a)$ and $G^b(f^b)$ denote the probability generating functions for distributions $P_a(k)$ and $P_b(k)$. Let $H^a(f^a)$ and $H^b(f^b)$ be the generating functions for excess degree distribution of network $A$ and $B$ respectively. The distribution of the number of links encountered by traversing a randomly chosen link (without including the randomly chosen link) is termed as the excess degree distribution. For networks generated by the configuration model \cite{Newman2010}, the excess degree distribution $Q(k)$, is given by $ \frac{1}{\langle k\rangle }(k+1)P(k+1)$, where $\langle k\rangle $ is the average degree and $P(k)$ is the degree distribution of the network. We assume that both the networks are generated using the configuration model.
   
 \section{Analytical Results \label{sec:Analysis}} 
We first review site and bond percolation on a single isolated network with degree distribution $P(k)$ and site (bond) occupation probability $p$, more details can be found in \cite{Newman2010}. In site percolation, each node is active with probability $p$ independent of other nodes, while in bond percolation each link is active with probability $p$. Thus, in site percolation $1-p$ fraction of nodes are removed, while in bond percolation $1-p$ fraction of links are removed. The connected component remaining after the node (link) removal is termed as the giant connected component (GCC). If $p$ is sufficiently low a GCC may not exist. Let $S$ denote the fraction of nodes in the GCC. Let $f$ be the probability that a randomly chosen link does not lead to the GCC. Therefore, for site percolation
\begin{align*}
S_{site} =  p\left(1 - \sum_{k=0}^{\infty}f^kP(k) \right)= p\left( 1 - G(f)\right)
\end{align*}   
and for bond percolation $S_{bond} =   1 - G(f)$, where $G(f)$ is the probability generating function of the degree distribution. A link does not lead to GCC if it is inactive, or if it is active and the node at the other end of the link does not belong to the GCC. Assuming that the network is generated by the configuration model, $f$ can be written as,
\begin{align}
f &= 1-p + \frac{p}{\langle k \rangle}\sum_{k=0}^{\infty}(k+1)P(k+1)f^{k} \nonumber \\
 f&=  1-p + pH(f) \label{eqn:bondpercolation}
\end{align}
where $H(f)$ is the generating function of the excess degree distribution $Q(k), \ Q(k) = \sum_{k=0}\frac{(k+1)P(k+1)}{\langle k \rangle} $. 
The fixed point of equation (\ref{eqn:bondpercolation}) can be calculated using the following iterative process \cite{Newman2010}.
\begin{align*}
f(i+1) = 1-p + p H(f(i))
\end{align*}  
Where $f(1), f(2),.., f(i)$ are the iterations. 
\par
The size of the GCC, (proportion of nodes in the GCC) is non zero if the solution of the fixed point equation is less than unity, i.e., $f<1$. This can happen if and only if $p>p_c$, where 
\begin{align*}
p_c = \frac{1}{H^{'}(1)} =  \frac{\langle k \rangle}{\langle k^2 \rangle - \langle k \rangle} 
\end{align*}
and $\langle k^2 \rangle$ is the second moment of the degree distribution. $p_c$ is also known as the critical threshold bond (or site) percolation probability.
\par
 The site and bond percolation process for percolation probabilities $p_{site}$ and $p_{bond}$ is summarized using the following equations
\begin{align*}
  S_{site}&= p_{site}(1 - G(f_{site})), \text{ where }  \\  
  f_{site}& = 1 - p_{site} + p_{site} H(f_{site}) \\
     S_{bond} &= p_{bond}(1 - G(f_{bond})), \text{ where } \\
      f_{bond} &= 1 - p_{bond} + p_{bond}H(f_{bond}) 
\end{align*} 
In a system consisting of interacting antagonistic and dependent networks, the size of the GCC in both networks may change with time. Let $x_n$ and $y_n$ be the fraction of active links in network $A$ and fraction of active nodes in $B$ respectively, at time step $n$. Similar to the single network case, we define $S_n^a$, $S_n^b$  and $f_n^a, \ f_n^b$ for networks $A$ and $B$ at time step $n$.  In each step, network $A$ can change, causing network $B$ to change in the same step. Note that bond percolation occurs in $A$, while site percolation occurs in $B$.
\par
 Since we have assumed that $N_a = N_b = N$, the probability that a node in $B$ is attacked ($\frac{rN_a}{N_b}$) becomes $r$. Now, assume that all the nodes in both networks are part of a large connected component. Compromised nodes launching a DDoS attack on nodes in $B$ initiates a site percolation process in $B$ with site occupation probability $y_1 = 1-r$.  This may result in fragmentation of network $B$ which then induces a bond percolation process on $A$, since links in $A$ are dependent on nodes in $B$. The changes in GCC in $A$ and $B$ are described by the following sequence.
 \par
Initially we assume that all nodes in $A$ are part of the giant connected component ($x_1 = 1$).
\begin{align*}
 S_1^a = 1 - G^a(f_1^a) , \text{ where }  f_1^a = H^a(f_1^a)
\end{align*}
 $r$ proportion of nodes in network $A$ launch an attack on network $B$, which induces a site percolation process on $B$. The proportion of nodes in $B$ which are attacked is $rS_1^a$. The site percolation probability is the proportion of nodes which are \emph{not} attacked by $A$, which is $1- rS_1^a$.  The resulting size of the GCC in network $B$ ($S_1^b$) is :   
\begin{align*}
  S_1^b &= y_1(1 - G^b(f_1^b)), \text{ where }   \\
   f_1^b &= 1 - y_1 + y_1H^b(f_1^b), \ y_1 = 1-rS_1^a 
   \end{align*}
   Since $q_a$ proportion of links in network $A$ depend on nodes in $B$, fragmentation of network $B$ induces a bond percolation process in $A$. Thus, $q_a(1-S_1^b)$ proportion of links stop functioning. The bond percolation probability is the proportion of functioning links in $A$, which is given by $1 - q_a(1-S_1^b)$. The resulting size of the GCC  at time step $2$ ($S_2^a$) in $A$ is given by: 
   \begin{align*} 
      S_2^a&= 1 - G^a(f_2^a) , \text{ where} \\
      f_2^a&= 1 - x_2 + x_2H^a(f_2^a), \ x_2= 1-q_a\left(1-S_1^b\right) 
    \end{align*}
     The bond percolation process on $A$ may result in fragmentation of the network, causing the attacking nodes to fail. Thus, the new site percolation probability is $1-rS_2^a$. The size of the GCC at time step $2$ due to the site percolation process is given by
    \begin{align*}
    S_2^b &= y_2(1 - G^b(f_2^b)), \text{ where } \\
     f_2^b &= 1 - y_2 + y_2H^b(f_2^b), \ y_2 = 1-rS_2^a  
   \end{align*}
   This process can continue for $n = 1,2,...$. Thus it can be written as
   \begin{align}
       S_n^a&= 1 - G^a(f_n^a), \text{ where } \nonumber \\ 
     f_n^a &= 1 - x_n + x_nH^a(f_n^a), \ x_n =  1-q_a\left(1-S_{n-1}^a\right) \nonumber \\
      S_n^b &= y_n(1 - G^b(f_n^b)), \text{ where }\nonumber \\
      f_n^b& = 1 - y_n + y_nH^b(f_n^b), \ y_n = 1-rS_n^a, \ \label{eqn:originalSystem} 
\end{align}
At equilibrium $\\ f^a=f_n^a = f_{n-1}^a , \ f^b = f_n^b = f_{n-1}^b \\$ and $ S^a=S_n^a = S_{n-1}^a , \ S^b = S_n^b = S_{n-1}^b \\$ Thus we obtain :
\begin{align}
S^a &= 1 - G^a(f^a) \nonumber \\
 f^a&= q_a\bigg(G^b(f^b) +r(1-G^a(f^a)) (1-G^b(f^b)) \bigg) \ +\nonumber \\ 
& H^a(f^a)\biggr [1-q_a\bigg(G^b(f^b) +r(1-G^a(f^a) )(1-G^b(f^b) )\bigg)\biggr] \label{eqn:fa}\\
 S^b &= (1 - r (1 - G^a(f^a)))(1 - G^b(f^b)) \nonumber \\
 f^b & = r(1-G^a(f^a)) + (1- r(1-G^a(f^a)) )H^b(f^b)  \label{eqn:fb}   
\end{align}
 \begin{figure}
 \includegraphics[width = 0.45\textwidth]{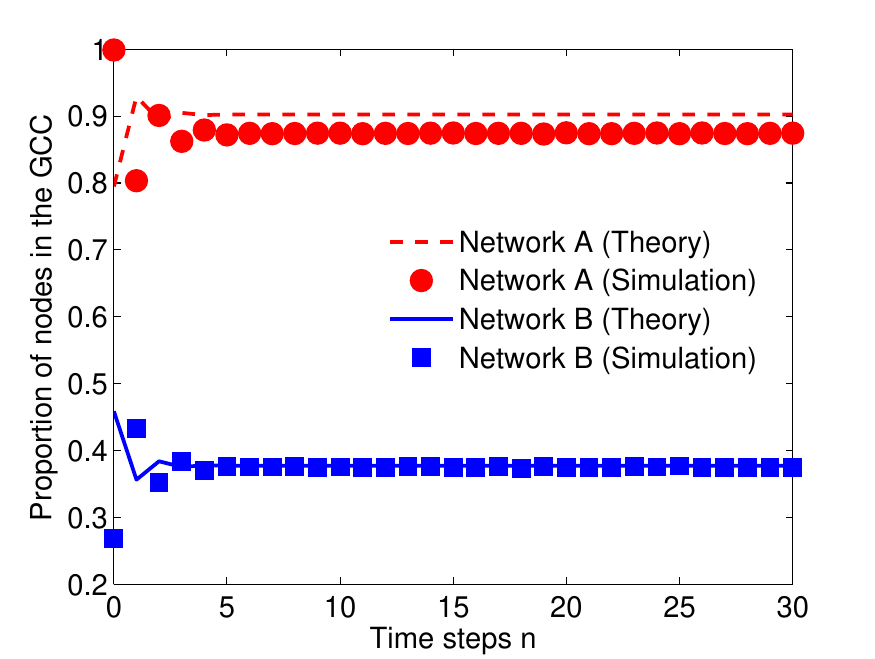}
 \caption{(Color online) Simulation results. Parameters: Erdos Renyi networks with mean degree $4.1$, $q_a = 0.7, \ r = 0.6$.}
 \label{fig:Simulations}
 \end{figure}
The above theoretical calculations are verified using extensive Monte Carlo simulations. The entire process described in the Model section, Sec. \ref{sec:Model} was simulated in a computer using the Java programming language, and the results were compared with the results obtained by numerically  evaluating system (\ref{eqn:originalSystem}). Fig. \ref{fig:Simulations} shows a good agreement between the theoretical predictions and the simulation results.
\subsection{Analysis of the Equilibrium Points}
We now check the conditions required for the boundary ($f^a,f^b \ \in \ \{0,1\}$) and non-boundary points ($f^a,f^b \in (0,1)$)  to be equilibrium points.  
\subsubsection{Complete collapse of both the networks}
If both networks do not percolate then $f^a = 1$ and $f^b = 1$. Substituting $f^b=1$ in equation (\ref{eqn:fa}) and $f^a = 1$ in equation (\ref{eqn:fb}) we obtain:
\begin{align*}
f^a = q_a + (1-q_a) H^a(f^a), \  f^b = H^b(f^b)
\end{align*}  
If $f^a=1$ and $f^b=1$ are the smallest fixed point solutions of the above equations, then the percolation probability must be less than the threshold ($p<p_c$) for both networks, i.e.,
\begin{align}
 q_a > 1-\frac{\langle k_a\rangle }{\langle k_a^2\rangle  - \langle k_a\rangle }   \text{ and }  \frac{\langle k_b^2\rangle  - \langle k_b\rangle }{\langle k_b\rangle } < 1 \label{eqn:complcollapse}
\end{align}  
The condition for network $B$ does not depend on $q_a$ and $r$, hence this corresponds to a scenario, where to begin with, network $B$ does not contain a GCC \cite{Durrett2007}. Thus, the complete fragmentation of both networks is possible if and only if network $B$ is completely fragmented from the very beginning. Since we assume that at the beginning of the process network $B$ contains a giant connected component (and thus is not fragmented), this scenario is not possible.  
 
 \subsubsection{Complete collapse of network $A$ }
 In this scenario $f^a = 1$ and either $f^b=0$ or $ f^b \in (0,1)$. If $f^b=0$, substituting $f^b=0$ in equation (\ref{eqn:fa}) we obtain 
 \par
 {\small
 \begin{align*}
 f^a = q_ar(1-G^a(f^a)) + \bigg[1-q_ar(1-G^a(f^a)) \bigg]H^a(f^a)
 \end{align*}
 }
 \par
 The condition for the non existence of GCC, i.e. $f^a=1$, is given by: 
    \begin{align*}
    \frac{\langle k_a^2\rangle  - \langle k_a\rangle }{\langle k_a\rangle } < 1
    \end{align*}
    The derivation of this condition can be found in the appendix. This condition implies that network $A$ does not contain a GCC from the very beginning. Since we assume that both networks are fully connected at the start, this scenario is not possible.
    \par
 In the other case: $f^b \in (0,1)$ and $f^a=1$, substituting $f^a = 1$ in equation (\ref{eqn:fb}) we obtain $f^b = H^b(f^b)$.  Thus, the conditions required for this point to be an equilibrium point are:
 \par
 {\small
 \begin{align}
 1 - q_aG^b(f^b) < \frac{\langle k_a\rangle }{\langle k_a^2\rangle  - \langle k_a\rangle }, \ \frac{\langle k_b^2\rangle  - \langle k_b\rangle }{\langle k_b\rangle } > 1  \label{eqn:facondition}
 \end{align}
 }
 \par
The derivation of the first condition can be found in the appendix. The second condition corresponds to $p>p_c$ for network $B$ (bond percolation probability must be greater than the critical threshold probability).
\par
 Consider an isolated network with distribution same as that of network $A$ which undergoes a \emph{random} attack. Let $q$ be the probability that a node is attacked in the isolated network. Assume that $q =1- \frac{\langle k\rangle }{\langle k^2\rangle -\langle k\rangle }$, i.e, GCC does not exist. Now, if $q$ is equal to $q_a$, then $1-q_aG^b(f^b) > 1-q$, i.e., the network with antagonistic and dependent interactions does not collapse. Thus, a GCC exists in network $A$, while it does not exist in the isolated network. In other words, networks with dependent antagonistic interactions are more robust against \emph{random} attacks than isolated networks.
 \par
 The intuition behind this mathematical result is as follows. When network $A$ launches a random attack on network $B$, the failure of nodes in network $B$ result in failure of links in network $A$. This results in reduction of the attack probability, since lesser number of nodes in $A$  attack $B$. Now, $q_a$ is the fraction of links in $A$ that depend on $B$, it is also the probability that a link in $A$ fails due to the collapse of network $B$. Due to antagonistic and dependent interactions the effective probability that a link in network $A$ collapses becomes $q_aG^b(f^b)$ which is less than $q_a$ since $G^b(f^b)<1$. Such an automatic reduction of link failure probability doesn't occur in isolated networks. Hence networks with dependent and antagonistic interactions are more robust against random attacks compared to isolated networks.  
  
\subsubsection{Complete collapse of network $B$}
Complete collapse of network $B$ is possible when $f^b=1$ and $f^a = 0$ or $0<f^a<1$.  Substituting $f^b=1$ in equation (\ref{eqn:fa}) we get
\begin{align*}
f^a = q_a + (1-q_a)H^a(f^a)    
\end{align*}
Clearly, $f^a$ cannot be zero since $q_a>0$. 
\par
The condition required for $0<f^a<1$ can be obtained by substituting $f^b=1$ in equation (\ref{eqn:fa}).  $f^b=1$ is possible if and only if a giant connected component does not exist in network $B$. This translates to:
\begin{align}
q^a &<1- \frac{\langle k_a\rangle }{\langle k_a^2\rangle  - \langle k_a\rangle } \nonumber \\
1-r(1-G^a(f^a)) &< \frac{\langle k_b\rangle }{\langle k_b^2\rangle -\langle k_b\rangle }  \label{eqn:fbcondition}
\end{align}
Since $1-G^a(f^a) < 1$, similar to the network $A$ case, it can be shown that this system is more robust against \emph{random} attacks compared to the isolated network case.
\par
   This can also be explained intuitive as follows. Now $r$ is the proportion of nodes in $A$ that launch an attack on network $B$.   The failure of nodes in network $B$, due to the attack, result in failure of links in network $A$ because a proportion of links in $B$ depend on nodes in $A$. This results in reduction of the attack probability, since lesser number of nodes in $A$  attack $B$. This is reflected in the equation: the attack probability is no longer $r$, but $r(1-G^a(f^a))$ which is less than $r$. Such a reduction does not occur in isolated networks since there is no negative feedback which can cause reduction in the random attack probability.
 
\subsubsection{Both networks percolate}
This is possible only when:  $f^a,f^b \in (0,1)$ since points with $f^a=0$ or $f^b=0$ cannot be equilibrium points. This is because $f^a=0$ or $f^b=0$ requires either of the conditions to be satisfied $q_a=0$ or $r=0$. 
       \begin{figure}
       \includegraphics[width = 80mm]{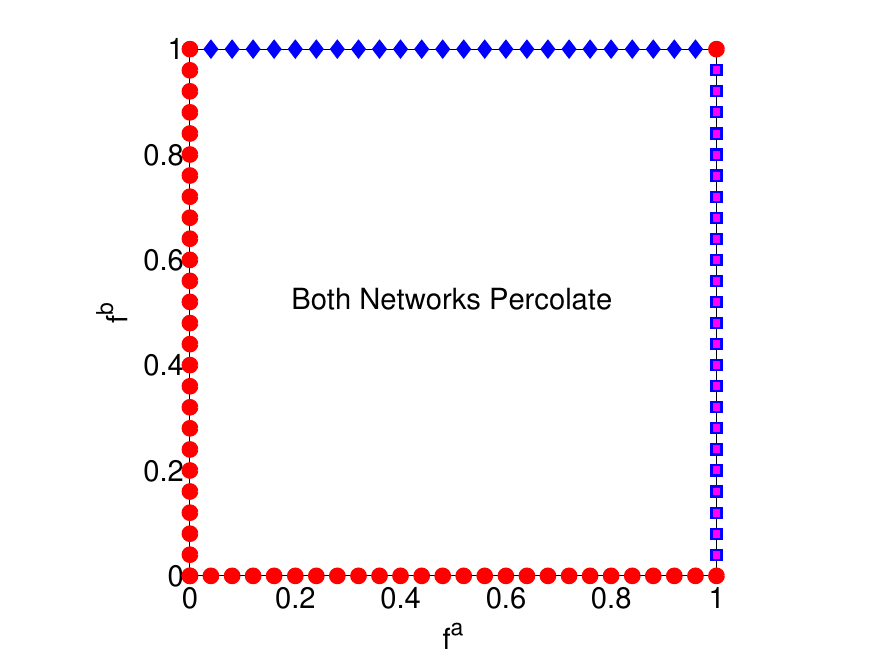}
       \caption{(Color Online) Feasible equilibrium points. Red circles : infeasible, magenta squares: $A$ collapses and $B$ percolates, blue diamonds: $A$ percolates and $B$ collapses, white unshaded region: both networks percolate    }
       \label{fig:FeasibleRegions}
       \end{figure}

 The conditions on $q_a$ and $r$ required for $f^a,f^b \in (0,1)$ can be obtained by numerically solving equations (\ref{eqn:fa}) and (\ref{eqn:fb}). Fig. \ref{fig:FeasibleRegions} pictorially illustrates the feasibility of equilibrium points.
 
\subsection{Stability Analysis}
       \begin{figure}
       \includegraphics[width = 80mm]{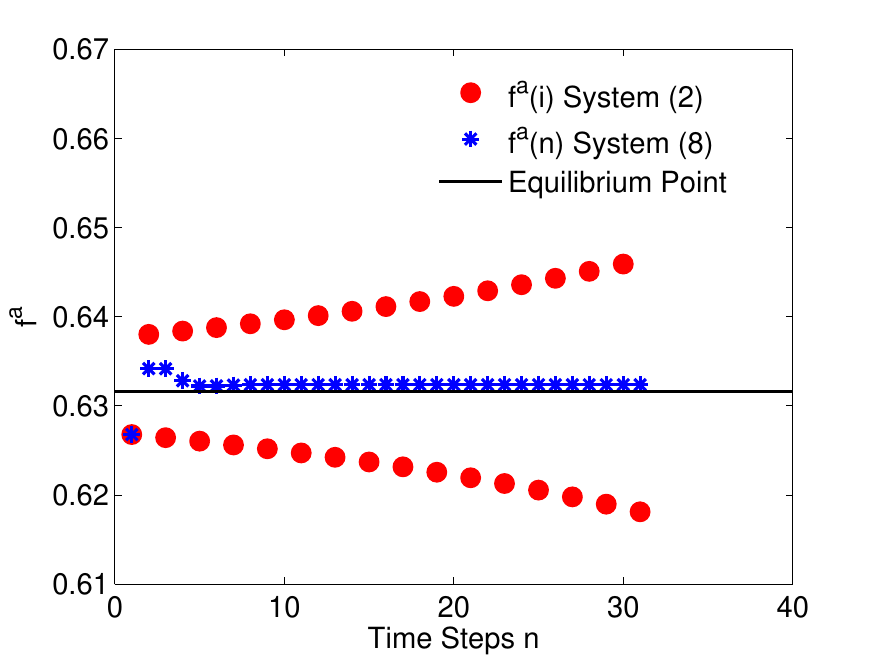}
       \caption{(Color Online) The $f^a$ sequence in system (\ref{eqn:originalSystem}) and (\ref{eqn:iterative}). Parameters: $r=0.78, \ q_a = 0.79$, Erdos-Renyi networks with average degree $4.1$ }
       \label{fig:sequence}
       \end{figure}
We now analyze the stability of the equilibrium points $f^{a^*}=1,f^{b^*} \in (0,1)$; $f^{a^*} \in (0,1),f^{b^*}=1$ and $f^{a^*},f^{b^*} \in (0,1)$.  Since system (\ref{eqn:originalSystem}) is complicated, it is difficult to determine the conditions under which the sequence ($f_1^a,f_2^a,...,f_n^a,...$) generated by it converges.  To tackle this problem, we introduce a new system with the same equilibrium points as (\ref{eqn:originalSystem}), but which generates a sequence which is bounded by the sequence generated by  (\ref{eqn:originalSystem}). Thus, if the new system is unstable then so is system (\ref{eqn:originalSystem}). The new system does not have any physical interpretation, it is constructed only for calculating the stability.  
\par
 The evolution of this new system is described as follows
 \begin{align}
 \hat{f}^a(i) &= U(\hat{f}^a(i-1),\hat{f}^b(i-1)) \nonumber \\
 \hat{f}^b(i) &= V(\hat{f}^a(i-1),\hat{f}^b(i-1)) \label{eqn:iterative}
 \end{align}
 where {\small 
\begin{align*}
& U(\hat{f}^a,\hat{f}^b) = q_a\bigg(G^b(\hat{f}^b) +((1-G^a(\hat{f}^a)) (1-G^b(\hat{f}^b)) \bigg) \ +\nonumber \\ 
& H^a(\hat{f}^a)\biggr [1-q_a\bigg(G^b(\hat{f}^b) +r(1-G^a(\hat{f}^a) )(1-G^b(\hat{f}^b) )\bigg)\biggr] \\
& V(\hat{f}^a,\hat{f}^b)  =  r(1-G^a(\hat{f}^a)) + [1- r(1-G^a(\hat{f}^a)) ]H^b(\hat{f}^b) 
\end{align*}
}
 Where $i$ is the iteration step, and $\hat{f}^a(i)$ is the value of $\hat{f}^a$ at the $i^{th}$ iteration. The equilibrium points of (\ref{eqn:iterative}) can be obtained by equating $\hat{f}^a(i)$ and $\hat{f}^b(i)$ with $\hat{f}^a(i-1)$ and $\hat{f}^b(i-1)$. The equilibrium point turns out to be equal to that of (\ref{eqn:originalSystem}). i.e., $\hat{f}^a = f^a$ and $ \hat{f}^b = f^b$. 
 \par
 System (\ref{eqn:iterative}) is not the same as system (\ref{eqn:originalSystem}), but they share the same equilibrium points. In (\ref{eqn:originalSystem}), $f_n^a$ and $f_n^b$ are fixed points for each time step $n$, while $\hat{f}^a(i), \hat{f}^b(i)$ in system (\ref{eqn:iterative}) are not fixed points. At each time step $n$ the GCC of networks $A$ and $B$ is calculated by evaluating the fixed points, and they then become an input to the next stage. 
\par
If the equilibrium point of a dynamical system is unstable, then under a small perturbation, the system diverges away from the equilibrium point. It can be show that, if $\hat{f}^a(i), \ \hat{f}^b(i)$ in (\ref{eqn:iterative}) diverge under perturbation, then $f_n^a,\ f_n^b$, in (\ref{eqn:originalSystem}), also diverge under perturbation. Thus, if the equilibrium point of the above system $(f^{a^*},f^{b^*})$ is unstable then the corresponding equilibrium point of system (\ref{eqn:originalSystem}), $(S^{a^{*}}, \ S^{b^{*}} )$, is also unstable. 
\par
To see this, consider a bond (or site) percolation process on a network, with bond (site) percolation probability $p$. As mentioned in the beginning of this section, the size of the GCC is given by $S_{bond} =   1 - G(f^*)$ where $f^*$ is the fixed point of the following equation
 \begin{align*}
 f = g(p,f) = 1-p + pH(f) 
 \end{align*}  
 Assume that $p$ is perturbed by $\Delta p$, $\Delta p > 0$, i.e., $p^{'} = p + \Delta p$. Since $\frac{\partial g}{\partial p} \leq 0$, the new fixed point, $f^{'*}$, is less than $f^*$, the old fixed point. Let $f(1) = g(p^{'},f^*)$; we term this as the first iteration $(i=1)$. Since the fixed point equation is increasing in $f$, i.e., $\frac{\partial g}{\partial f} > 0 $, and $f^{'*} < f^*$, we have $g(p^{'},f^{'*}) < g(p^{'},f^{*})$, i.e., $f^{'*}<f(1)$. For a positive $\Delta p$, the new fixed point is always lesser than the first iteration. Similarly, for $\Delta p < 0$, the fixed point is always greater than the first iteration. 
 \par
 The points $\hat{f}^a(i),\hat{f}^b(i)$ in system (\ref{eqn:iterative}) are first iteration points for ($i=1$), i.e., if $f^{b^*}$ is perturbed then a new value $\hat{f}^a(i+1)$ is calculated incorporating the perturbation.The points $\hat{f}^a(i),\hat{f}^b(i)$ are not fixed points, while the points $f_n^a,f_n^b$ in system (\ref{eqn:originalSystem}) are fixed points. Since fixed points are always greater than first iterations, if the perturbation is negative, and lesser if the perturbation is positive, the sequence generated by system (\ref{eqn:iterative}) is  bounded on both sides by the sequence generated by system (\ref{eqn:originalSystem}). 
 \par
 This is illustrated in Fig. \ref{fig:sequence}. Both systems are perturbed (negative perturbation of $f^{a^*}$). The sign of subsequent perturbations alternate between positive and negative. As seen in the figure, system (\ref{eqn:originalSystem}) bounds system (\ref{eqn:iterative}) from both sides.  Hence, if (\ref{eqn:iterative}) diverges under perturbation, the corresponding fixed points of (\ref{eqn:originalSystem}) also diverge under perturbation.      
 \par
 However, the opposite may not be true, i.e., stability of system (\ref{eqn:iterative}) may not imply stability of (\ref{eqn:originalSystem}). Thus, the stability of system (\ref{eqn:iterative}) is a necessary condition for $S^a_n$ and $S^b_n$ in (\ref{eqn:originalSystem}) to be stable, or in other words if system (\ref{eqn:iterative}) is unstable at the equilibrium point, then system (\ref{eqn:originalSystem}) is also unstable at the same equilibrium point.
\par
 We analyze the stability of system (\ref{eqn:iterative}) using linear stability analysis. The equilibrium point $(f^{a^*},f^{b^*})$ is stable if and only if the magnitude of each eigen value of the Jacobian matrix $J$ is less than one.
 \par
\begin{align*}
J = 
\begin{bmatrix}
\frac{\partial U}{\partial \hat{f}^a}\bigg |_{f^{a^*},f^{b^*}} & \frac{\partial U}{\partial \hat{f}^b}\bigg |_{f^{a^*},f^{b^*}} \\
\frac{\partial V}{\partial \hat{f}^a}\bigg |_{f^{a^*},f^{b^*}} & \frac{\partial V}{\partial \hat{f}^b}\bigg |_{f^{a^*},f^{b^*}}
\end{bmatrix}
\end{align*}

\begin{align}
&\left( \lambda - \frac{\partial U}{\partial \hat{f}^a} \right) \left( \lambda - \frac{\partial V}{\partial \hat{f}^b}   \right)  
- \frac{\partial U}{\partial \hat{f}^b}\frac{\partial V}{\partial \hat{f}^a}\bigg |_{\hat{f}^a=f^{a^*},\hat{f}^b=f^{b^*}} = 0 \label{eqn:stability}
\end{align} 

In the following sections, we calculate the necessary conditions for stability of equilibrium points $(f^{a^*}=1,0<f^{b^*}<1)$ and $(0<f^{a^*}<1,f^{b^*}=0)$. Stability conditions for the equilibrium point  $f^{a^*},f^{b^*} \in (0,1)$ can be calculated numerically by computing the roots of equation (\ref{eqn:stability}).

\subsubsection{Stability of the equilibrium point $(f^{a^*}=1, 0<f^{b^*}<1)$}
After substituting $\hat{f}^a = 1$ and $\hat{f}^b = f^{b^*}$ in equation (\ref{eqn:stability}), the eigen values are:
\begin{align}
\lambda_1 &= (1-q_aG^b(f^{b^*}))\left( \frac{\langle k_a^2\rangle -\langle k_a\rangle }{\langle k_a\rangle } \right)  \nonumber \\
\lambda_2 &= H^{b'}(f^{b^*})  \label{eqn:stabilitycondn1}
\end{align}
From condition (\ref{eqn:facondition}), $\lambda_1 < 1$. Thus, the point is stable if and only if $H^{b'}(f^{b^*}) < 1$. The points $q_a^*$ and $r^*$ which satisfy this condition can be computed numerically by evaluating conditions (\ref{eqn:stabilitycondn1}) and (\ref{eqn:facondition}). 

\subsubsection{Stability of the equilibrium point $(0<f^{a^*}<1, f^{b^*}=1)$}
The roots of the characteristic equation after substituting $\hat{f}^a= f^{a^*}$ and $\hat{f}^b=1$ are:
\begin{align}
\lambda_1 &= (1-q_a)H^{a'}(f^{a^*}) \label{eqn:stabilitycondn2}\\
\lambda_2 &= [1-r(1-G^a(f^{a^*}))]\left(\frac{\langle k_b^2\rangle -\langle k_b\rangle }{\langle k_b\rangle } \right) \nonumber
\end{align}
From condition (\ref{eqn:fbcondition}), $\lambda_2<1$. Thus the point is stable if and only if $(1-q_a)H^{a'}(f^{a^*}) <1$. The region can be computed by numerically evaluating conditions (\ref{eqn:stabilitycondn2})  and (\ref{eqn:fbcondition}).

\subsection{Nature of the phase transition}
 \begin{figure}
 \includegraphics[width = 0.45\textwidth]{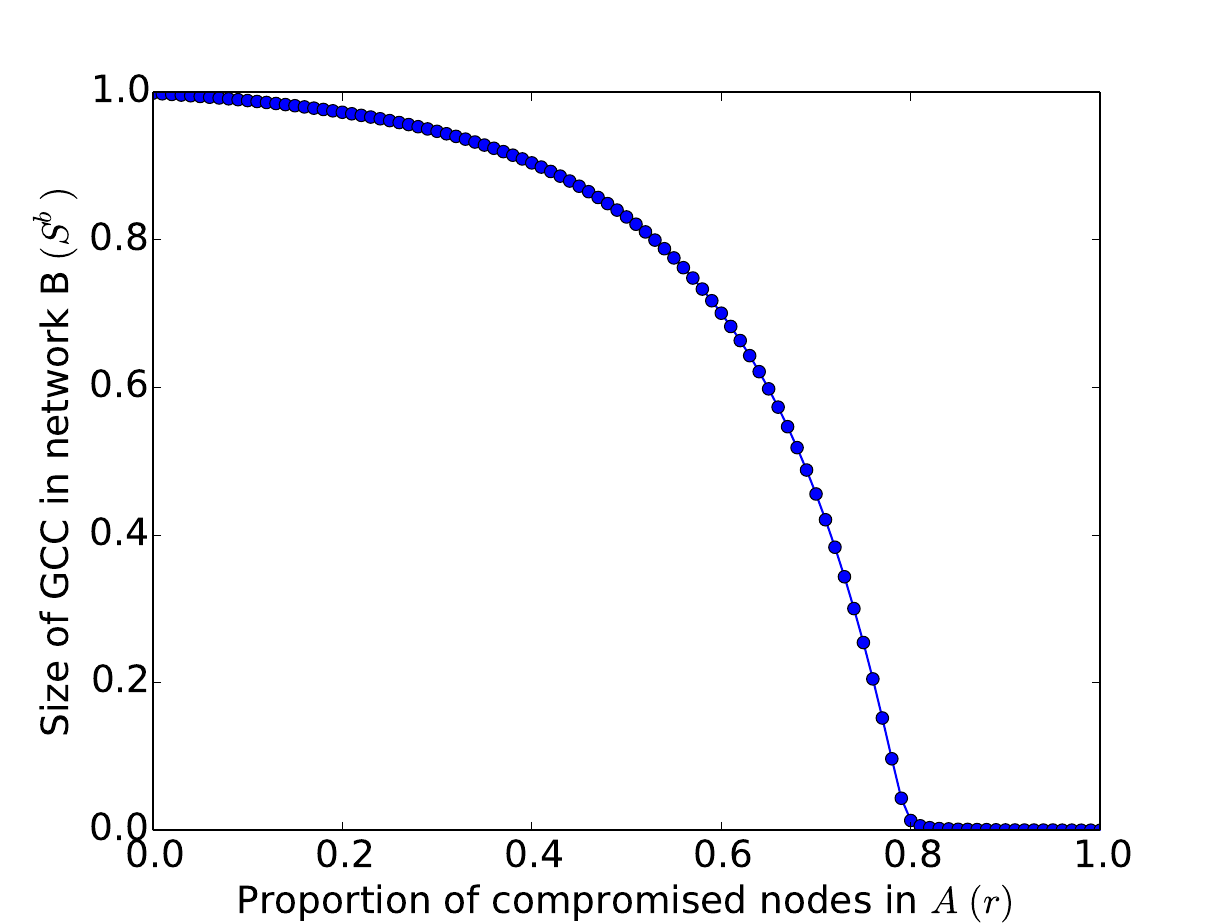}
 \caption{(Color online) Continuous phase transition. Parameters: Erdos-Renyi networks with mean degree $4.1$, $q_a = 0.3$. At $r=0.8$, $f^{b^*}=1$}
 \label{fig:PhaseTransition}
 \end{figure}

Fig. \ref{fig:PhaseTransition} shows a phase transition in network $B$ when both networks are Erdos-Renyi and have the same mean degree. Thus a phase transition may  occur near the equilibrium points $(f^{a^*}=1,0<f^{b^*}<1)$ and $(0<f^{a^*}<1,f^{b^*}=1)$ if they exist and are stable.  
Using the Implicit Function Theorem we show that the phase transition, in both the cases, is continuous in $q_a$ and $r$. 
Let
{\small
\begin{align*}
& g(f^a,f^b) =q_a\bigg(G^b(f^b) +r(1-G^a(f^a)) (1-G^b(f^b)) \bigg) \ +\nonumber \\ 
& H^a(f^a)\biggr [1-q_a\bigg(G^b(f^b) +r(1-G^a(f^a) )(1-G^b(f^b) )\bigg)\biggr] - f^a \\
& h(f^a,f^b)  =  r(1-G^a(f^a)) + [1- r(1-G^a(f^a)) ]H^b(f^b)  - f^b
\end{align*}
}
\par
Equations (\ref{eqn:fa}) and (\ref{eqn:fb}) can be written as $g(f^a,f^b) = 0$ and $h(f^a,f^b) = 0$.
  
\subsubsection{Phase transition at $(f^{a^*}=1, 0<f^{b^*}<1)$ }   
A phase transition is possible only if the equilibrium point is stable, and hence we assume that the conditions for stability of $(f^{a^*}=1, 0<f^{b^*}<1)$  are satisfied.
We first show that the derivative of $f^a$ with respect to $r$ and $q_a$ near $f^{a^*}=1$ exists using the Implicit Function Theorem.  Since $g(f^a,r,q_a)$ is continuously differentiable, according to the Implicit Function Theorem
\begin{align*}
\frac{df^a}{dr} = -\frac{\partial g}{\partial r} \bigg /  \frac{\partial g}{\partial f^a}
\end{align*} 

The derivative exists at $f^a = f^{a^*}$ if and only if $\frac{\partial g}{\partial f^a}\big |_{f^a=f^a*} \neq 0$. The same existence condition holds true for $\frac{df^a}{dq_a}$
{\small
\begin{align*}
\frac{\partial g}{\partial f^a} &= H^{a'}\bigg[1-q_a\bigg(G^b+r(1-G^a )(1-G^b) \bigg)\bigg] -1 \\
&+ (1-H^a)q_a\bigg(G^{b'} \frac{\partial f^b}{\partial f^a} ( 1- r(1-G^a)) - rH^a(1-G^b)  \bigg)
\end{align*}
}
where 
\begin{align*}
\frac{\partial f^b}{\partial f^a} &= -rG^{a'} + rG^{a'}H^{b}  + (1-r(1-G^{a}))H^{b'}\frac{\partial f^b}{\partial f^a}  
\end{align*}
At $f^a = f^{a^*} = 1$, 
\begin{align*}
\frac{\partial f^b}{\partial f^a}  = \frac{rG^{a'}(1)(1-H^{b}(f^b))}{1-H^{b'}(f^b)} < \infty
\end{align*}
This is because we have assumed that the point is stable, and hence it satisfies the condition $H^{b'} (f^b)< 1$.  Simplifying we get
\begin{align*}
\frac{\partial g}{\partial f^a}  = H^{a'}(1)(1-q_aG^{b}(f^b))  \neq 0
\end{align*}
This is because, since $f^b < 1$, $1-q_aG^{b}(f^b) \neq 0$ (because $p>p_c, \ p = 1-q_aG^{b}(f^b)$). Thus $f^a$ is differentiable and hence continuous with respect to $r$ and $q_a$ at $f^a = 1$. Since $S^a$ is a continuous (polynomial) function of $f^a$, $S^a$ is continuous in $q_a$ and $r$. Thus, the phase transition is continuous.

\subsubsection{Phase transition at $(0<f^{a^*}<1, f^{b^*}=1)$}
Assuming the equilibrium point is stable, we use the Implicit Function Theorem on $h(f^a,f^b)$ to show that the derivative of $f^b$ with respect to $q_a$ and $r$ exists. By the Implicit Function Theorem

\begin{align*}
\frac{df^b}{dr} = -\frac{\partial h}{\partial r} \bigg / \frac{\partial h}{\partial f^b}
\end{align*} 
\par
{\small
\begin{align*}
\frac{\partial h}{\partial f^b} = -rG^{a'}\frac{\partial f^a}{\partial f^b}(1-H^b) + [1-r(1-G^a)]H^{b'}
\end{align*}
}
At $f^b = f^{b^*} = 1$, 
\begin{align*}
\frac{\partial f^a}{\partial f^b} = \frac{q_a(1-H^a(f^a))G^{b'}(1)(1-r(1-G^a(f^a)))}{1 - H^{a'}(f^a)(1-q_a)} 
\end{align*}
 As the point is stable, $(1-q_a)H^{a'}(f^a)<1$, and therefore $\frac{\partial f^a}{\partial f^b}  < \infty$. Hence at $f^b = f^{b^*}=1$
\begin{align*}
\frac{\partial h}{\partial f^b} = [1-r(1-G^a(f^a))]\left(\frac{\langle k_b^2\rangle -\langle k_b\rangle }{\langle k_b\rangle }\right)  \neq 0
\end{align*}
This is because,  $1-G^a(f^a) < 1 $ since $f^a > 0$.
\vspace{.1in}
\par
Thus, $f^b$ is differentiable and hence continuous with respect to $q_a$ and $r$ at $f^b=1$. This implies that $S^b$ is also continuous at the phase transition point with respect to $q_a$ and $r$. 
 
 \section{Numerical Results \label{sec:Results}}
       \begin{figure}
       \includegraphics[width = 0.45\textwidth]{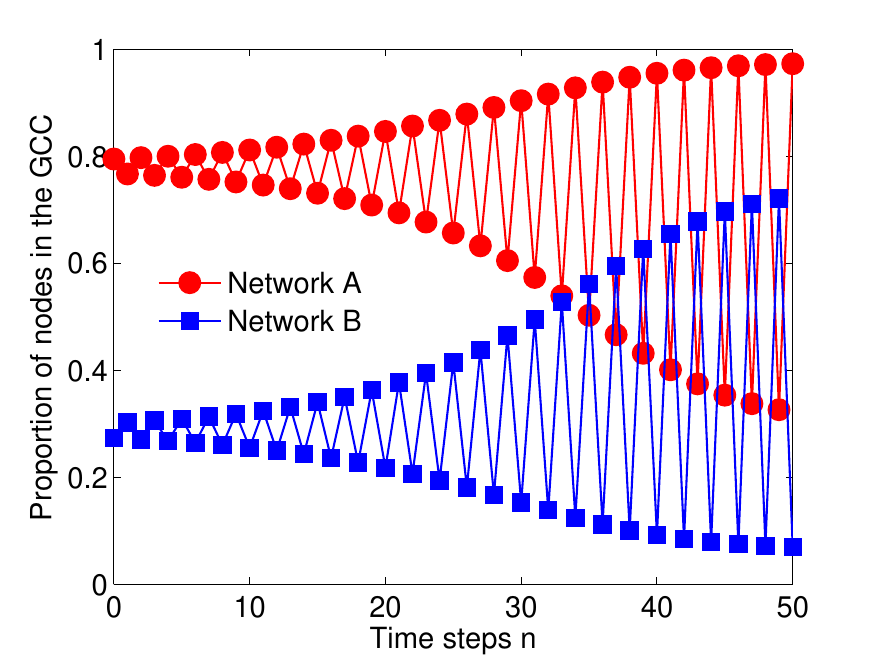}
       \caption{(Color online) Oscillation of GCC in both networks. Parameters: Erdos-Renyi networks with mean degree $4.1$, $q_a=0.81, \ r = 0.78$.}
       \label{fig:oscillations}
       \end{figure}
  \begin{figure*}[t]
      \centering
      \begin{subfigure}[]{80mm}
      
      \includegraphics[width = 80mm]{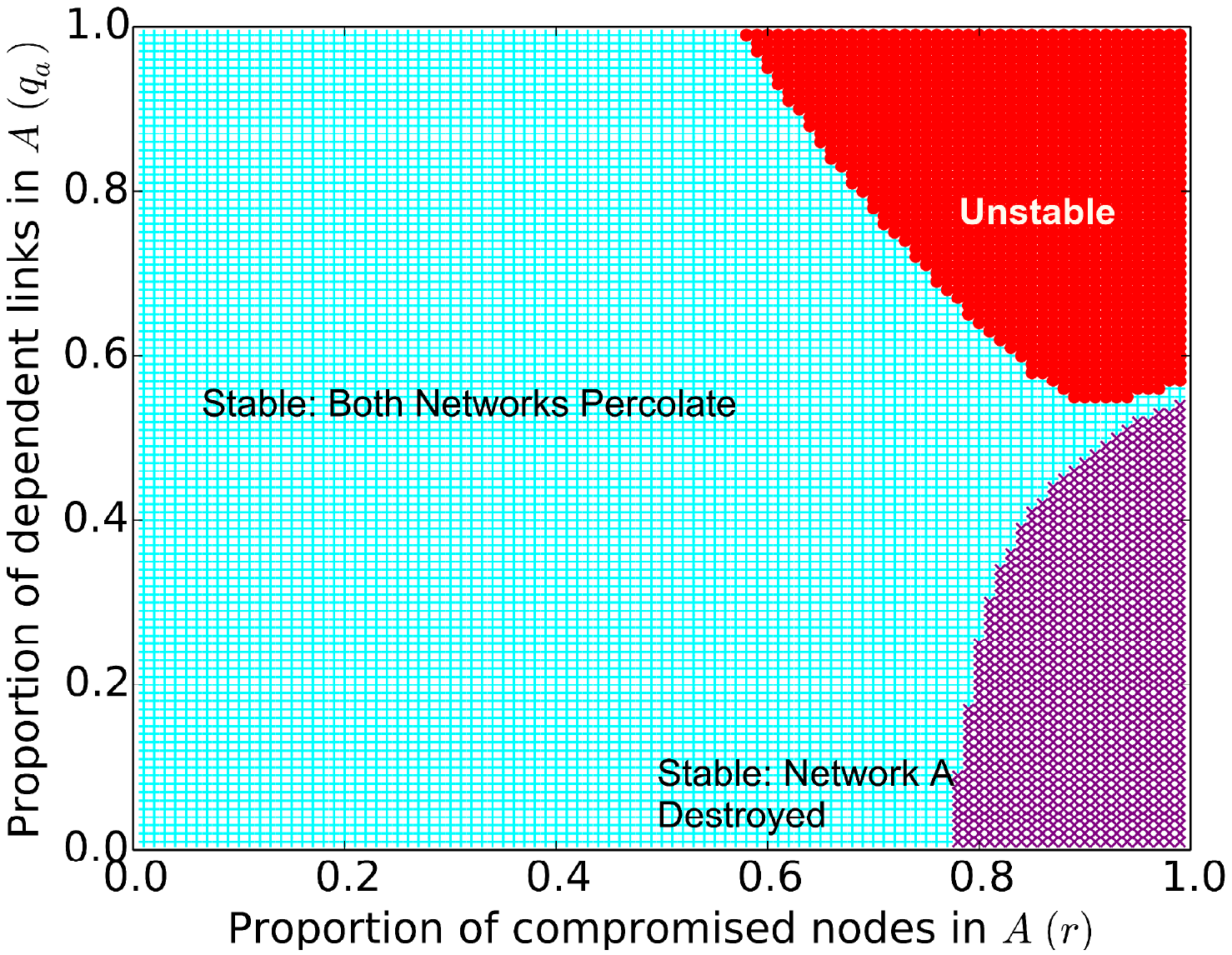}
                      \caption{}
      					\label{fig:ERboth}
      \end{subfigure}
     \hspace{4pt}
      \begin{subfigure}[]{80mm}
      \hspace{4pt}
      \includegraphics[width = 80mm]{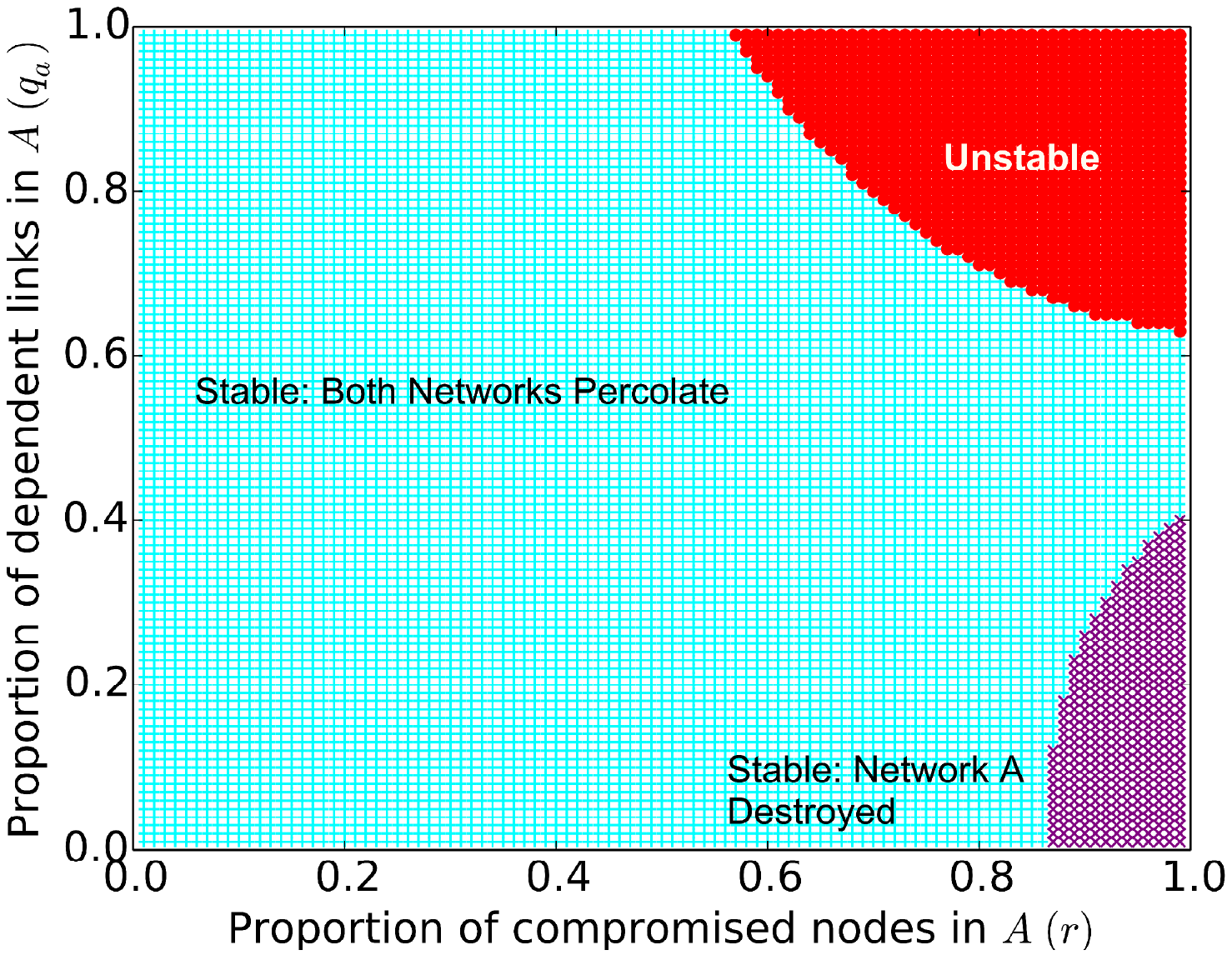}
                      \caption{}
                      \label{fig:SFboth}
       \end{subfigure}
  
        \begin{subfigure}[]{80mm}
              
              \includegraphics[width = 80mm]{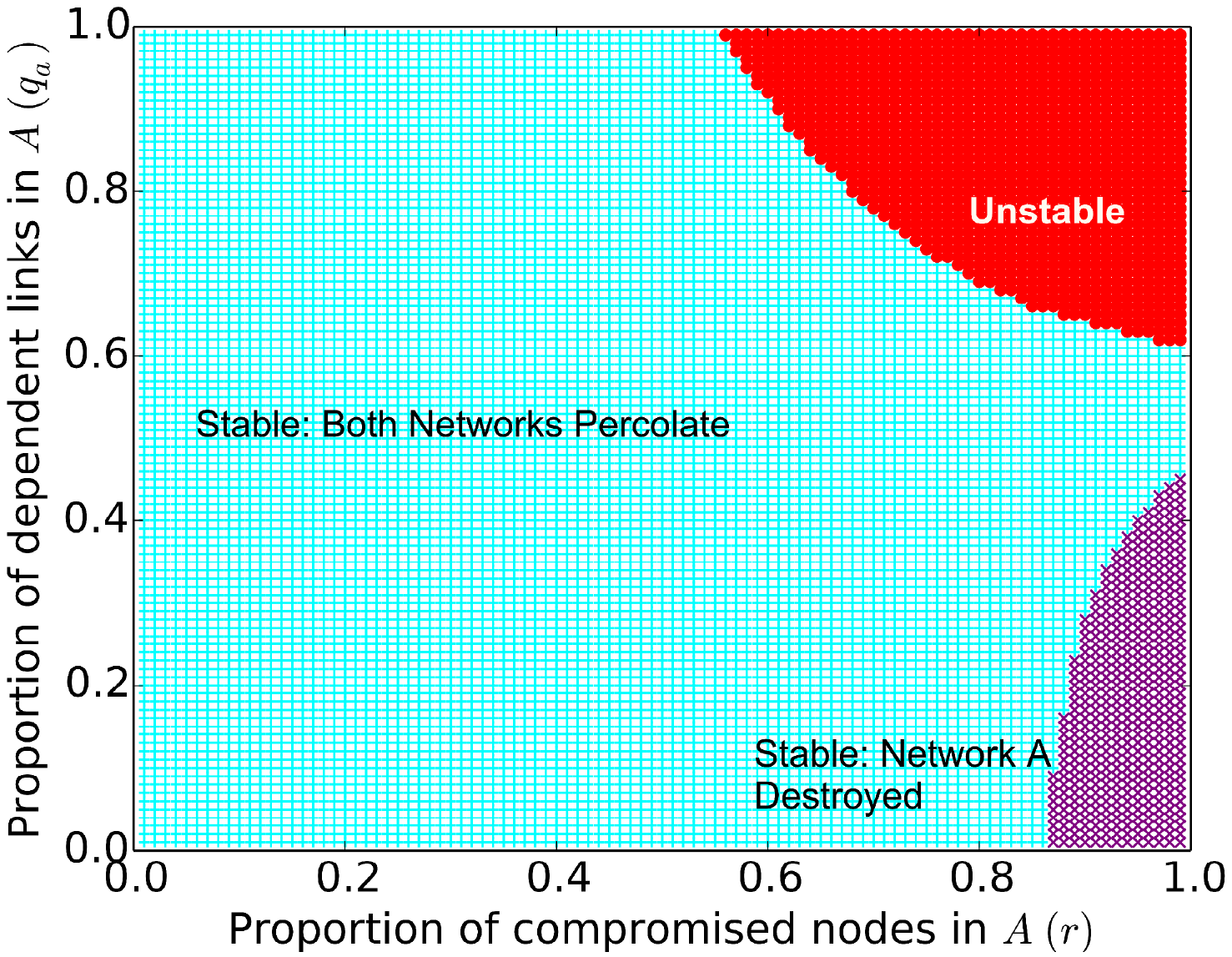}
                              \caption{}
              					\label{fig:ERA}
              \end{subfigure}
               \hspace{4pt}
              \begin{subfigure}[]{80mm}
              
              \includegraphics[width = 80mm]{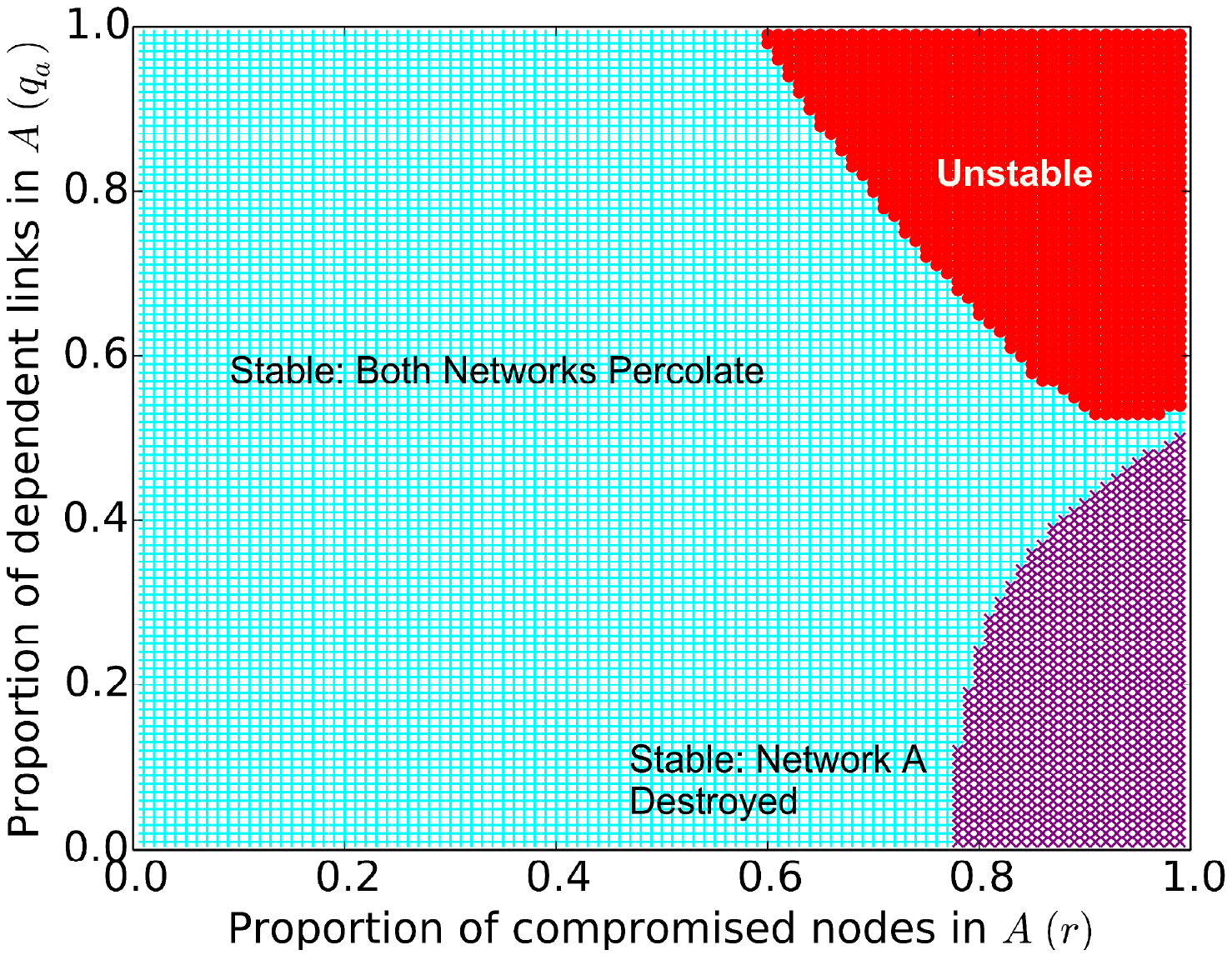}
                              \caption{}
                              \label{fig:SFA}
               \end{subfigure}
       \caption{(Color online) Regions in Erdos-Renyi (ER) and Scale Free (SF) networks. (a) both are ER networks with average degree $4.1$. (b) both are SF  networks with the same mean degree $4.1$ and exponent of $2.8$, (c) $A$ is an ER network, $B$ is an SF network, both with average degree $4.1$, (d) $A$ is a SF  network while $B$ is an ER network both with the same mean degree $4.1$. Purple crosses represent the region where $A$ is destroyed, light blue squares designate the region where both networks percolate, and red circles represents the unstable region. }
      \end{figure*}

Using numerical computations we find that the solution of system (\ref{eqn:originalSystem}) may display periodic oscillations. This is illustrated in Fig. \ref{fig:oscillations} for the case when both networks are Erdos-Renyi. However, these oscillations do not occur for all parameter values; in fact, for low $r$ and $q_a$ the system is stable, as also seen in Fig. \ref{fig:PhaseTransition}.  Thus, the parameter space, $r,q_a \ \in (0,1] $, can be divided into four regions.
\begin{enumerate}
\item Network $A$ destroyed, $B$ percolates, no oscillations.
\item Network $A$ percolates, $B$ destroyed, no oscillations.
\item Both networks percolate, no oscillations.
\item System oscillates.
\end{enumerate}
 \par
   Fig. \ref{fig:ERboth} shows the regions when both networks are Erdos-Renyi, while Fig. \ref{fig:SFboth} shows the regions when both networks are scale free.  Since scale free networks are more robust against random attacks, region $2$ is smaller in Fig. \ref{fig:SFboth} than in Fig. \ref{fig:ERboth}.  Fig. \ref{fig:ERA} shows the regions when $A$ is an Erdos Renyi network and $B$ is a scale free network, both with the same mean degree, while  Fig. \ref{fig:SFA} displays the regions when $A$ is scale free network and $B$ is an Erdos-Renyi network.  The figures show that region $2$ shrinks only when $B$ is a scale free network. This suggests that given that both networks have the same mean degree, the structure of network $A$ does not influence the robustness of network $B$. Thus, a scale free power distribution network, irrespective of the attacking network structure, is more robust than an Erdos-Renyi power distribution network, against random attack.  In all the cases region $1$ is absent, i.e., there does not exist parameter values required for network $A$ to collapse.
  
\section{Discussion \label{sec:Discussion}} 
 In this article we considered a system consisting of two networks exhibiting dependent and antagonistic interactions, i.e., a network whose nodes are antagonistic towards the nodes of the other network, while a portion of its links are dependent on the other network. Such a situation may arise when a botnet launches DDoS attacks against SCADA  switches which control the power stations of a power distribution network. Failure of power stations may cause the links connecting the botnet to fail resulting in cessation of the attack. 
 \par
 Our analysis showed that unlike interdependent networks \cite{Gao2012} and interacting networks exhibiting antagonistic interactions \cite{Zhao2013,Zhao2014}, which can display first order phase transition, the phase transitions observed in the system considered  here are continuous. Also, unlike interacting antagonistic networks \cite{Zhao2013,Zhao2014}, we do not find the existence of bistability in the solutions. In comparison with an isolated network, interdependent networks are more fragile against random attack \cite{Huang2011,Buldyrev2011}, while the system discussed here is more robust against \emph{random} attacks.  This shows that such systems behave very differently from interdependent and interacting antagonistic networks.  Also, for Erdos-Renyi and scale free networks, we do not find a region in the parameter space where the antagonistic network collapses. In other words, the attacking network does not collapse even though its links are dependent on the victim network.
 \par
 Numerical calculations revealed a region in the parameter space where the giant connected components of both the networks start oscillating. Outside this region the giant connected components of both the networks are stable. Numerical results suggest that oscillations occur when the antagonism and the dependence are very high, i.e., a large proportion of nodes in the antagonistic network launch an attack on the victim network, while a large proportion of links in the antagonistic network are dependent on the nodes of the victim network. 
 \par      
 In this article we studied a system where bond percolation process occurs on one network while site percolation happens on the other. A system where bond or site percolation happens on both networks with antagonistic and dependent interactions is likely to exhibit similar results. We believe that this article provides valuable insights on the percolation behavior in systems consisting of networks exhibiting antagonistic and dependent interactions. Such studies are key for developing an understanding of real world interconnected and interdependent systems.

 \section*{Appendix \label{sec: appendix}}
 The complete collapse of network $A$ is possible if and only if $f^a=1$ and either $f^b=0$ or $0<f^b<1$.  We aim to calculate the conditions required for the emergence of the GCC in network $A$. We first consider the second case: $0<f^b<1$. If $f^b  \in (0,1)$, $f^a$ must satisfy 
 \begin{align*}
  f^a &= 1-T(f^a) + T(f^a) H^a(f^a)
 \end{align*}
 where 
 \begin{align*}
 T(f^a) = 1-q_a\bigg( G^b(f^b) + r(1-G^a(f^a))(1-G^b(f^b))\bigg)
 \end{align*}
 The necessary and sufficient condition for the existence of a giant connected component is $f^a < 1$.
  Let, $u \ \in \ [0,1]$, be given by
 \begin{align*}
 u = 1 + \frac{f^a-1}{T(f^a)}
 \end{align*}
  Hence, $f^a = 1 + (u-1)T(f^a)$. Substituting this in the above equation of $f^a$ we obtain
  \begin{align*}
  u &= H^a(1+(u-1)T(f^a))  \\
  S^a &= 1- G^a(1+(u-1)T(f^a))
  \end{align*}
  This is equivalent to bond percolation with bond occupation probability $T(u)$. Proceeding in a manner detailed in Ref. \cite{Newman2002}, let $H_0^a(u;T(u))$ be the generating function for the size of a cluster in $A$  starting from a randomly chosen node, while $H_1^a(u;T(u))$ be the size of cluster reached by following a randomly chosen edge.  For a large $N$, the clusters are tree like, which allows one to write
  \begin{align*}
  H_0^a(u;T(f^a)) &= uG^a(1+(H_0^a(u;T(f^a))-1)T(f^a)) \\
  H_1^a(u;T(f^a)) &= uH^a(1+(H_0^a(u;T(f^a))-1)T(f^a))
  \end{align*}
   The mean size of the cluster, $\langle s^a\rangle $, is given by.
   \begin{align*}
   \langle s^a\rangle  &= \frac{\partial }{\partial u}H_0^{a}(u;T(f^a)) \bigg |_{u=1} \\ 
   &= 1 + G^{a'}(1)H_1^{a'}(1;T(1))T^{'}(1)
   \end{align*}
When $u=1$, $T(f^a) = T(1)$ because $f^a = 1$ if $u=1$. 
   Now,
   \par
   {\small
   \begin{align*}
   H_1^{a'}(1;T(1))T^{'}(1)  &= 1 + H^{a'}(1)[ T(1)H_1^{a'}(1;T(1)) + T^{'}(1) ] \\
   H_1^{a'}(1;T(1))T^{'}(1)  &=  \frac{1+H^{a'}(1)T^{'}(1)}{1-H^{a'}(1)T(1)}
   \end{align*}
   }
   \par
   The above equation diverges when $T(1) = T_c = \frac{1}{H^{a'}(1)}$,  hence for $T(1) > T_c$, we have a giant connected component or $f^a<1$. The condition for $f^a=1$ is given by 
   \begin{align*}
   1 - q_aG^b(f^b) < \frac{\langle k_a\rangle }{\langle k_a^2\rangle  - \langle k_a\rangle }
   \end{align*}
   If $f^b=0$, then the above condition becomes
   \begin{align*}
   \frac{\langle k_a^2\rangle  - \langle k_a\rangle }{\langle k_a\rangle } < 1
   \end{align*}
   This condition implies that a giant connected component does not exist in $A$ at the very beginning. 
 \bibliography{DependentAnatagonistic}

\end{document}